\documentclass[conference]{IEEEtran}
\IEEEoverridecommandlockouts
\usepackage{cite}
\usepackage{amsmath,amssymb,amsfonts}
\usepackage{algorithmic}
\usepackage{graphicx}
\usepackage{textcomp}
\usepackage{booktabs}
\usepackage{xcolor}
\usepackage{etoolbox}
\usepackage{graphicx}
\usepackage{subcaption}
\makeatletter
\patchcmd{\@makecaption}
  {\scshape}
  {}
  {}
  {}
\makeatletter
\patchcmd{\@makecaption}
  {\\}
  {.\ }
  {}
  {}
\makeatother
\def\tablename{Table}

\renewcommand{\figurename}{Figure}

\def\BibTeX{{\rm B\kern-.05em{\sc i\kern-.025em b}\kern-.08em
    T\kern-.1667em\lower.7ex\hbox{E}\kern-.125emX}}
\begin{document}

\title{Editable-DeepSC: Cross-Modal Editable Semantic Communication Systems}

\author{
    \IEEEauthorblockN{
        Wenbo Yu\IEEEauthorrefmark{3}\IEEEauthorrefmark{1},
        Bin Chen\IEEEauthorrefmark{3}\IEEEauthorrefmark{2},
        Qinshan Zhang\IEEEauthorrefmark{4},
        and Shu-Tao Xia\IEEEauthorrefmark{4}
    }
    \IEEEauthorblockA{
        \IEEEauthorrefmark{3} School of Computer Science and Technology, Harbin Institute of Technology, Shenzhen \\
        \IEEEauthorrefmark{4} Tsinghua Shenzhen International Graduate School, Tsinghua University \\
        yuwenbo@stu.hit.edu.cn;
        chenbin2021@hit.edu.cn;
        qs-zhang21@mails.tsinghua.edu.cn;
        xiast@sz.tsinghua.edu.cn
    }
\thanks{\IEEEauthorrefmark{1} Wenbo Yu performed this work while pre-admitted to Tsinghua Shenzhen International Graduate School.}
\thanks{\IEEEauthorrefmark{2} Bin Chen is the Corresponding Author.}
}

\maketitle

\begin{abstract}
Different from data-oriented communication systems that primarily focus on how to accurately transmit every bit of data, task-oriented semantic communication systems only transmit the specific semantic information required by downstream tasks, strive to minimize the communication overhead and maintain competitive tasks execution performance in the presence of channel noise. However, it is worth noting that in many scenarios, the transmitted semantic information needs to be dynamically modified according to the users' preferences in a conversational and interactive way, which few existing works take into consideration. In this paper, we propose a novel cross-modal editable semantic communication system, named Editable-DeepSC, to tackle this challenge. By utilizing inversion methods based on StyleGAN priors, Editable-DeepSC takes cross-modal text-image pairs as the inputs and transmits the edited information of images based on textual instructions. Extensive numerical results demonstrate that our proposed Editable-DeepSC can achieve remarkable editing effects and transmission efficiency under the perturbations of channel noise, outperforming existing data-oriented communication methods.
\end{abstract}

\section{Introduction}
Shannon and Weaver \cite{shannon1948mathematical, weaver1953recent} first proposed the mathematical formulation of a general communication system and divided the goals of communication into three levels, i.e., transmission of the symbols, transmission of the semantic information behind the symbols, and effects of the semantic information transmission. In the past decades, most of the traditional communication systems were designed for the first level of goals, namely to reduce bit error rate (BER) or symbol error rate (SER) as much as possible. In traditional communications, well-known source coding methods (e.g., JPEG \cite{wallace1992jpeg}, Huffman \cite{huffman1952method}) efficiently collaborate with commonly used channel coding methods (e.g., RS \cite{reed1960polynomial}, LDPC \cite{gallager1962low}) and have obtained excellent data recovery effects through noisy channels.

Recently, with the rapid development of deep learning techniques, Deep Neural Networks (DNN) are increasingly applied in communication systems. By adopting an end-to-end training scheme, DNN integrate multiple physical layers and overcome many inherent drawbacks in conventional communication systems. Bourtsoulatze \emph{et al.} \cite{bourtsoulatze2019deep} proposed Deep Joint Source-Channel Coding (DeepJSCC), which does not suffer from the \emph{cliff effect} and exhibits a graceful performance decline curve as the channel SNR varies while the traditional methods behave in the opposite way. All these communication methods mentioned above, whether implementing DNN or not, can be classified as \emph{data-oriented} communication methods \cite{wallace1992jpeg, huffman1952method, reed1960polynomial, gallager1962low, bourtsoulatze2019deep, zhang2023adaptive}. \emph{Data-oriented} communications aim to completely recover the transmitted data at the receiver side, regardless of the users' usage about the data. But obviously, different downstream tasks require different types of semantic information from the original data, which should be encoded and transmitted with different importance. Reconstructing every part of the original data equally will undoubtedly waste the limited bandwidth.

Contrary to \emph{data-oriented} communications, semantic communications are often \emph{task-oriented}, only transferring the semantic information suitable for downstream tasks. As a result, \emph{task-oriented} communications can exploit much more of the scarce bandwidth to transmit task-related information, thus realizing better tasks execution effects. Xie \emph{et al.} \cite{xie2021deep, xie2021task} contributed several profound semantic communication works, concerning single-modal text translation tasks, multi-modal Visual Question Answering (VQA) tasks, etc. Weng \emph{et al.} \cite{weng2023deep} explored the semantic communication problems from the perspective of speech tasks. These \emph{task-oriented} communication methods dived down into different modalities or practical scenarios, achieving superior tasks execution performance compared with \emph{data-oriented} methods.

However, few existing methods take into consideration that the transmitted semantic information needs to be dynamically adjusted according to the users' requirements in many scenarios. For example, in many famous social platforms (e.g., Facebook, Instagram), before transmitting their own photos to the remote servers, users often wish to flexibly edit the original multimedia data according to their personal needs, such as adding smile to a face or altering the transparency of eyeglasses. Furthermore, such personal requirements can be fulfilled by providing dialogues to enable a more conversational and interactive experience. Handling the semantic communication problems under these widely applicable scenarios is surely of great significance.

\begin{figure}[t]
\centerline{\includegraphics[width=0.5\textwidth]{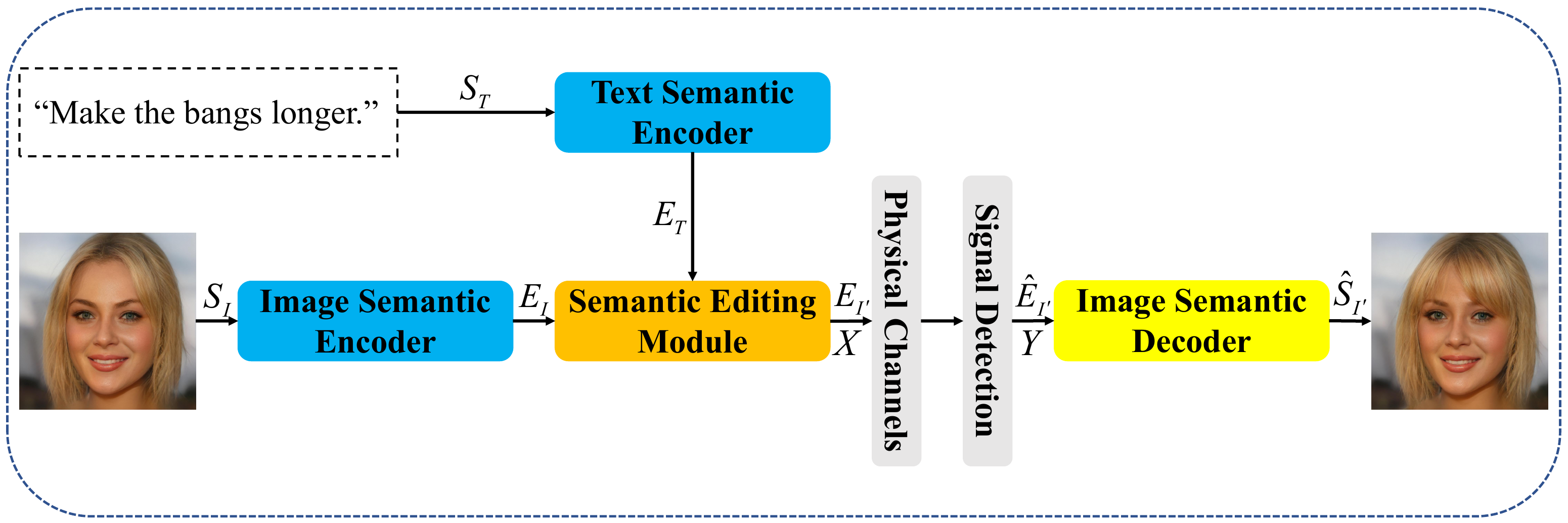}}
\caption{The overall architecture of our proposed \emph{Editable-DeepSC}. Our model mainly consists of the cross-modal codecs and the \emph{Semantic Editing Module}.}
\label{architecture}
\end{figure}

Therefore, we propose a novel cross-modal editable semantic communication system, i.e., \emph{Editable-DeepSC}, to fill this gap. As shown in Figure \ref{architecture}, \emph{Editable-DeepSC} leverages Generative Adversarial Networks (GAN) inversion methods \cite{goodfellow2020generative, xia2022gan, fang2023gifd, fang2024privacy} to encode the input images into the latent space based on StyleGAN \cite{karras2019style} priors. Meanwhile, the textual instructions are also encoded to guide the editing process. Since the image semantic information has been fully disentangled in the StyleGAN latent space, fine-grained editings can be realized through directly modifying the latent codes via the \emph{Semantic Editing Module}, even under extreme channel conditions.

\begin{figure*}[htbp]
\centerline{\includegraphics[width=2\columnwidth]{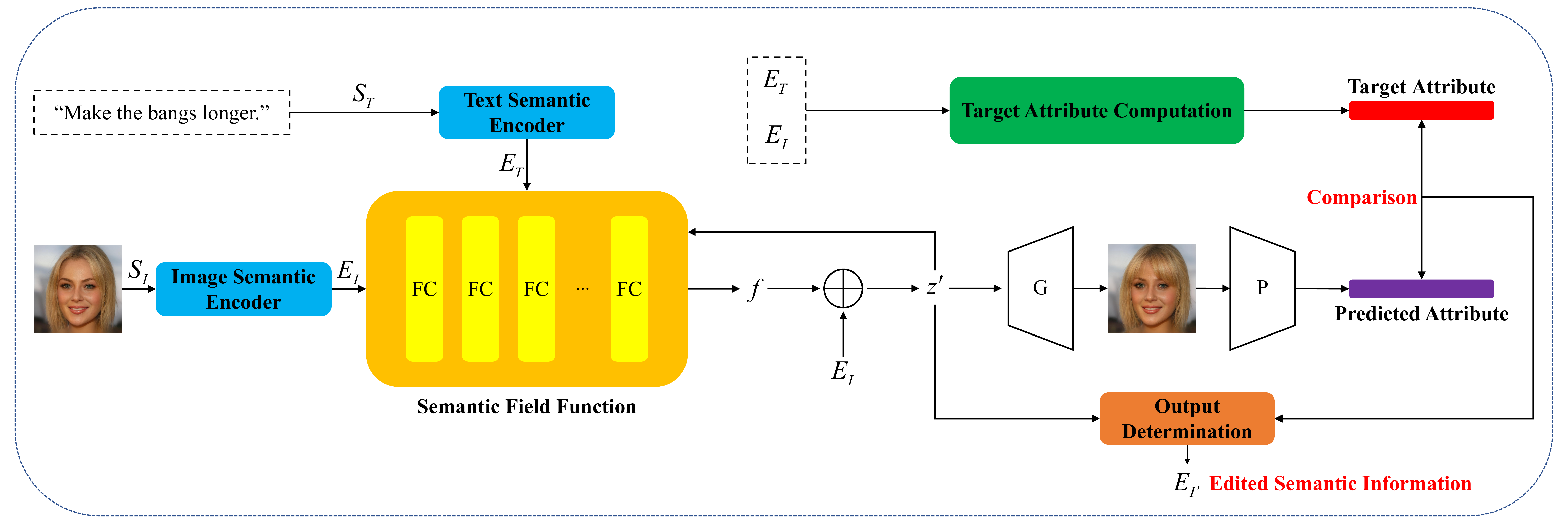}}
\caption{The working procedure of the \emph{Semantic Editing Module} in our proposed \emph{Editable-DeepSC}. The expected degree of the attribute, e.g., the length of the bangs, will be computed according to the given encodings $E_{I}$ and $E_{T}$. We model the \emph{Semantic Field Function} to exert minor modifications on $E_{I}$. By comparing whether the predicted attribute matches with the target attribute after each refinement on $E_{I}$, fine-grained editings can ultimately be realized.}
\label{semantic_editing_module}
\end{figure*}

\section{System Model}

We consider a cross-modal text-driven image editing scenario, which is illustrated in Figure \ref{architecture}. The input image $S_{I}$ is first transformed into a latent representation, i.e.,

\begin{equation}
    E_{I} = \mathcal{SE_{I}}(S_{I};\alpha),
\end{equation}

\noindent where $\mathcal{SE_{I}}(\cdot;\alpha)$ denotes the \emph{Image Semantic Encoder} with the parameters $\alpha$.

Similarly, the corresponding instructive sentence $S_{T}$ is also encoded by the \emph{Text Semantic Encoder} $\mathcal{SE_{T}}(\cdot;\beta)$ with the parameters $\beta$:

\begin{equation}
    E_{T} = \mathcal{SE_{T}}(S_{T};\beta).
\end{equation}

To acquire the edited latent code $E_{I^{\prime}}$, the multi-modal embeddings $E_{I}$ and $E_{T}$ are then sent into the \emph{Semantic Editing Module} $\mathcal{M}(\cdot, \cdot;\gamma)$ with the parameters $\gamma$:

\begin{equation}
    E_{I^{\prime}} = \mathcal{M}(E_{I}, E_{T};\gamma).
\end{equation}

The edited semantic information $E_{I^{\prime}}$ (also denoted as $X$) is then transmitted through the physical channels. We enforce an average transmission power constraint on $X$:

\begin{equation}
    \frac{1}{k} \cdot ||X||^{2}_{2} \leq P,
\end{equation}

\noindent where $k$ is the length of $X$ and $P$ is the power constraint.

Following the previous work \cite{bourtsoulatze2019deep} in this research field, the Channel Bandwidth Ratio (CBR) can be defined as:

\begin{equation}\label{CBR}
    \rho = \frac{k}{H \times W \times C},
\end{equation}

\noindent where $H$, $W$, and $C$ represent the image's height, width, and color channels. Smaller $\rho$ indicates better compression.

The transmitted signals are usually corrupted by the channel noise. Consequently, only the disrupted forms of the edited semantic information can be detected at the receiver side, i.e.,

\begin{equation}
    Y = h \ast X + N,
\end{equation}

\noindent where $Y$ (also denoted as $\hat{E}_{I^{\prime}}$) represents the received edited semantic information, $h$ denotes the channel coefficients, and $N$ denotes the Gaussian noise, whose elements are independent of each other and have the same mean and variance.

Finally, $\hat{E}_{I^{\prime}}$ can be mapped back to the vision domain so as to get the edited image anticipated by the sender, i.e.,

\begin{equation}
    \hat{S}_{I^{\prime}} = \mathcal{SD_{I}}(\hat{E}_{I^{\prime}};\theta),
\end{equation}

\noindent where $\mathcal{SD_{I}}(\cdot;\theta)$ indicates the \emph{Image Semantic Decoder} with the parameters $\theta$. Note that only the edited images are ultimately required, and thus the \emph{Text Semantic Decoder} is \emph{unnecessary} in our model.

\section{Methodology}

\subsection{Implementation Details of the Model}

Our model mainly consists of the cross-modal codecs and the \emph{Semantic Editing Module}. $E_I$ and $E_T$ are sent to the \emph{Semantic Editing Module}, which will iteratively update $E_I$ until the predicted degree output by the pre-trained attribute predictor matches with the expected degree.

\subsubsection{Image Semantic Encoder \& Decoder}

We leverage the GAN inversion methods based on StyleGAN priors to design $\mathcal{SE_{I}}(\cdot;\alpha)$. To be specific, given an input image $S_I$, a random vector $z$ is first initialized in the StyleGAN latent space. The random vector $z$ is then fed into the pre-trained StyleGAN Generator $G$ so as to generate an initial image $I_g=G(z)$. The distortion $J(z)$ between $S_I$ and $I_g$ is calculated as:

\begin{equation}
    J(z) = \lambda_{inv} \cdot MSE(S_I, I_g) + LPIPS(S_I, I_g).
\end{equation}

\noindent The distortion $J(z)$ is measured from both the pixel level and the perceptual level, balanced by the hyper-parameter $\lambda_{inv}$. In this paper, we adopt the MSE loss for the pixel level distortion and the LPIPS loss \cite{zhang2018unreasonable} for the perceptual level distortion. To obtain the latent code that perfectly suits the original image, we can minimize $J(z)$ and update $z$ by iteratively performing the gradient descent until convergence:

\begin{equation} \label{inversion}
    z_{t+1} = z_{t} - \eta \nabla_{z_{t}} J(z_{t}).
\end{equation}

\noindent As for the \emph{Image Semantic Decoder}, i.e., $\mathcal{SD_{I}}(\cdot;\theta)$, we send $\hat{E}_{I^{\prime}}$ to the StyleGAN Generator to recover the edited image.

\subsubsection{Text Semantic Encoder}

We utilize the Long Short-Term Memory (LSTM) \cite{hochreiter1997long} network to encode the textual instructions. The original texts are tokenized into words or subwords according to the dictionary before being sent to the LSTM network. The ultimate text encodings will contain the semantic information needed for editing, e.g., the attribute of interest, the direction of editing, and the degree of modification.

\subsubsection{Semantic Editing Module}

Figure \ref{semantic_editing_module} depicts how the \emph{Semantic Editing Module} comes into effect. The expected degree of the attribute, e.g., the length of the bangs, will first be computed according to the given encodings $E_I$ and $E_T$ by the \emph{Target Attribute Computation} unit. We also utilize fully connected networks to model the \emph{Semantic Field Function}, which can exert minor modifications on $E_{I}$ and obtain $z^{\prime}$. Since we leverage the inversion methods to extract vision features as shown in (\ref{inversion}), the semantic information in the StyleGAN latent space can be sufficiently disentangled, which means that fine-grained editings on the attribute of interest can be realized by iteratively shifting the latent code, without influencing other unrelated attributes. As shown in Figure \ref{semantic_editing_module}, the \emph{Semantic Field Function} takes $E_I$, $E_T$, and $z^{\prime}$ as the inputs $[E_I, E_T, z^{\prime}]$ to decide how to move $z^{\prime}$ at this round of iteration. After each round of movement, $z^{\prime}$ is delivered to the StyleGAN Generator $G$ to reconstruct an intermediate image, which will be further sent to the pre-trained attribute predictor $P$ in \cite{jiang2021talk} to check whether the anticipated requirement is satisfied. The predicted degree of the intermediate image calculated by $P$ will be compared with the expected degree. Once they match with each other, the iteration will stop and the \emph{Output Determination} unit will treat $z^{\prime}$ as the output $E_{I^{\prime}}$. Otherwise, the latent code will continue to iteratively make small movements until the target is reached or the maximum number of iterations is exceeded.

\subsection{Training Strategy}

The attributes of the image for editing, e.g., bangs, eyeglasses, smiling, are quantified as $[a_1, a_2, ..., a_i, ..., a_m]$, where the value of $a_i \; (1 \leq i \leq m)$ indicates the degree of each attribute. To train the \emph{Semantic Field Function} for the $i$-th attribute, we set the target degree of the $i$-th attribute to be $(a_i + 1)$ so that the model can learn how to perform fine-grained editings on each attribute. The predictor loss $\mathcal{L}_{pred}$ can be calculated by adopting the cross-entropy loss:

\begin{equation}
    \mathcal{L}_{pred} = - \sum_{i = 1}^{m} \sum_{j = 0}^{u} b_{ij} \cdot log(p_{ij}),
\end{equation}

\noindent where $b_{ij} \in \{0, 1\}$ is the one-hot representation of the expected attribute degrees, $p_{ij}$ is the output of the pre-trained attribute predictor $P$ on the edited image, and $u$ is the maximum value of $a_i$.

We also leverage the identity preservation loss $\mathcal{L}_{id}$ to better keep the original facial identity. We use ready-made facial recognition model to extract the features. The extracted features for facial recognition should be as similar as possible:

\begin{equation}
    \mathcal{L}_{id} = || F(I_a) - F(I_b) ||_{1},
\end{equation}

\noindent where $I_a, I_b$ are the images after and before the editing, and $F(\cdot)$ is the pre-trained face recognition model \cite{deng2019arcface}.

The discriminator loss $\mathcal{L}_{disc}$ is also adopted to ensure the fidelity on the generated images:

\begin{equation}
    \mathcal{L}_{disc} = - D(I_a),
\end{equation}

\noindent where $D(\cdot)$ is the output of the discriminator. Finally, the total loss can be calculated as:
\begin{equation}
    \mathcal{L} = \lambda_{pred} \cdot \mathcal{L}_{pred} + \lambda_{id} \cdot \mathcal{L}_{id} + \lambda_{disc} \cdot \mathcal{L}_{disc},
\end{equation}

\noindent where $\lambda_{pred}$, $\lambda_{id}$,  and $\lambda_{disc}$ are the weight factors. We utilize the total loss $\mathcal{L}$ to train the \emph{Semantic Editing Module} $\mathcal{M}(\cdot, \cdot;\gamma)$ through the noisy channels.

\subsection{Evaluation Metrics}

\begin{figure*}[tbp]
    \centering
    \begin{subfigure}{0.32\textwidth}
        \includegraphics[width=\linewidth]{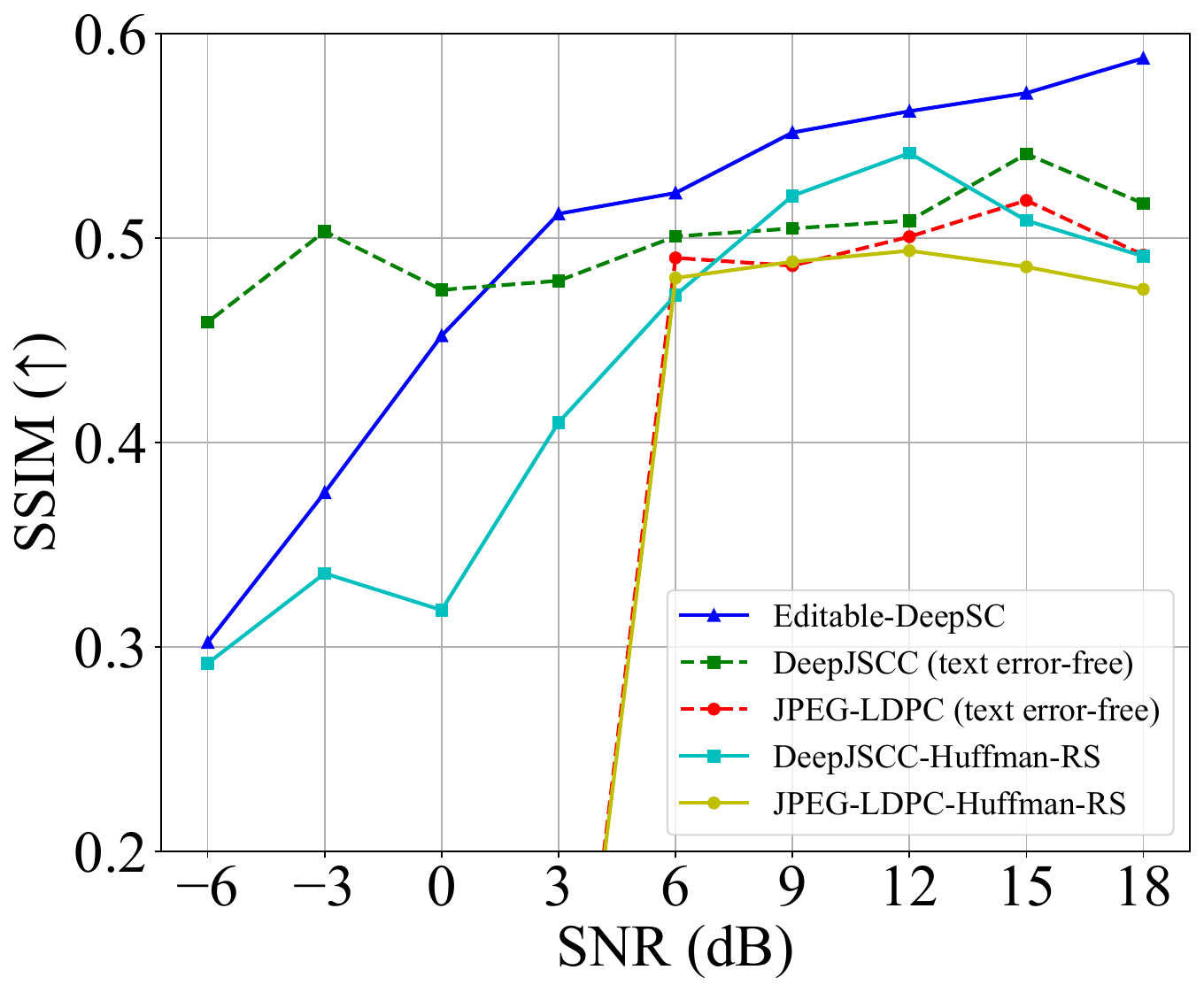}
        \label{ssim}
    \end{subfigure}
    \begin{subfigure}{0.32\textwidth}
        \includegraphics[width=\linewidth]{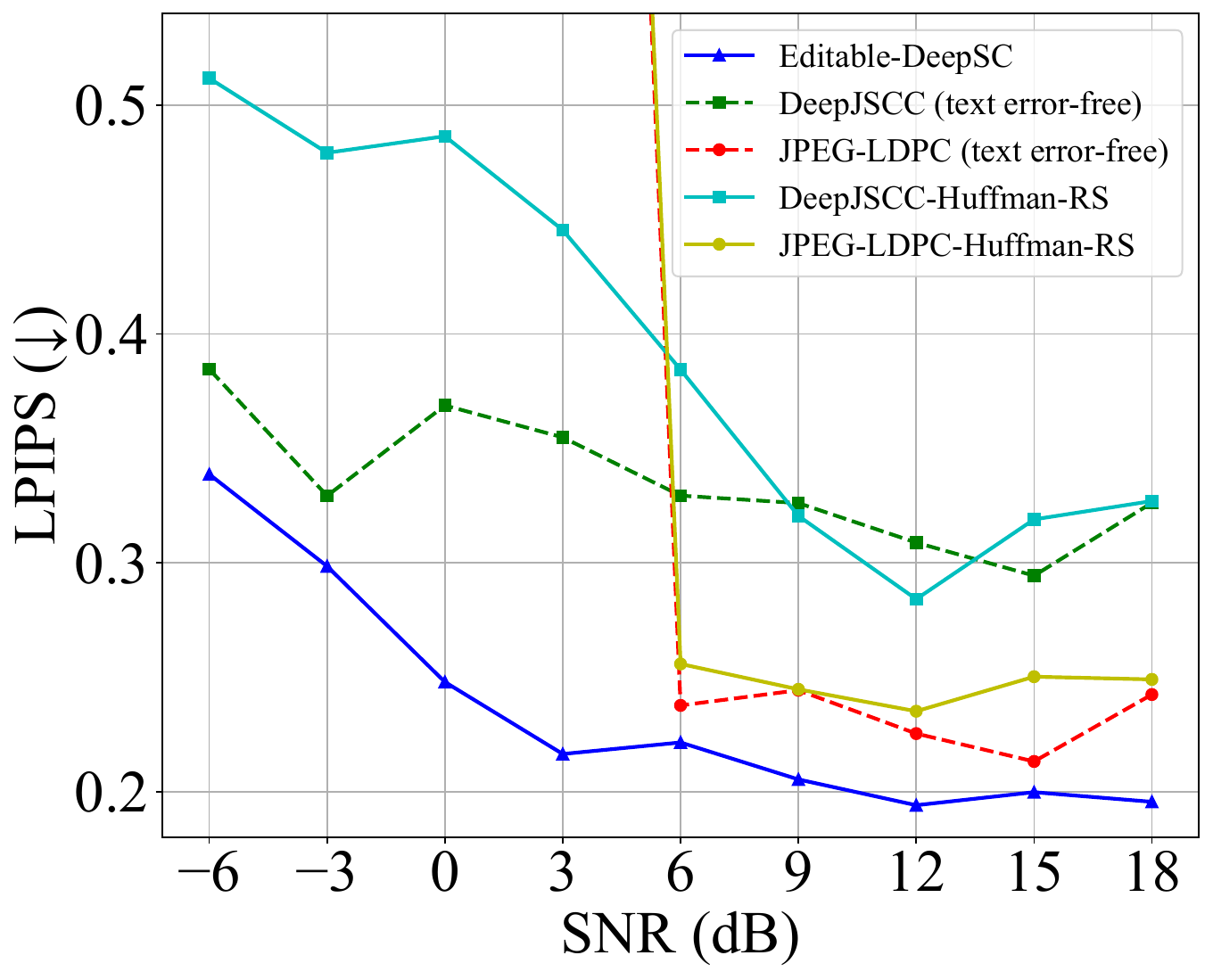}
        \label{lpips}
    \end{subfigure}
    \begin{subfigure}{0.32\textwidth}
        \includegraphics[width=\linewidth]{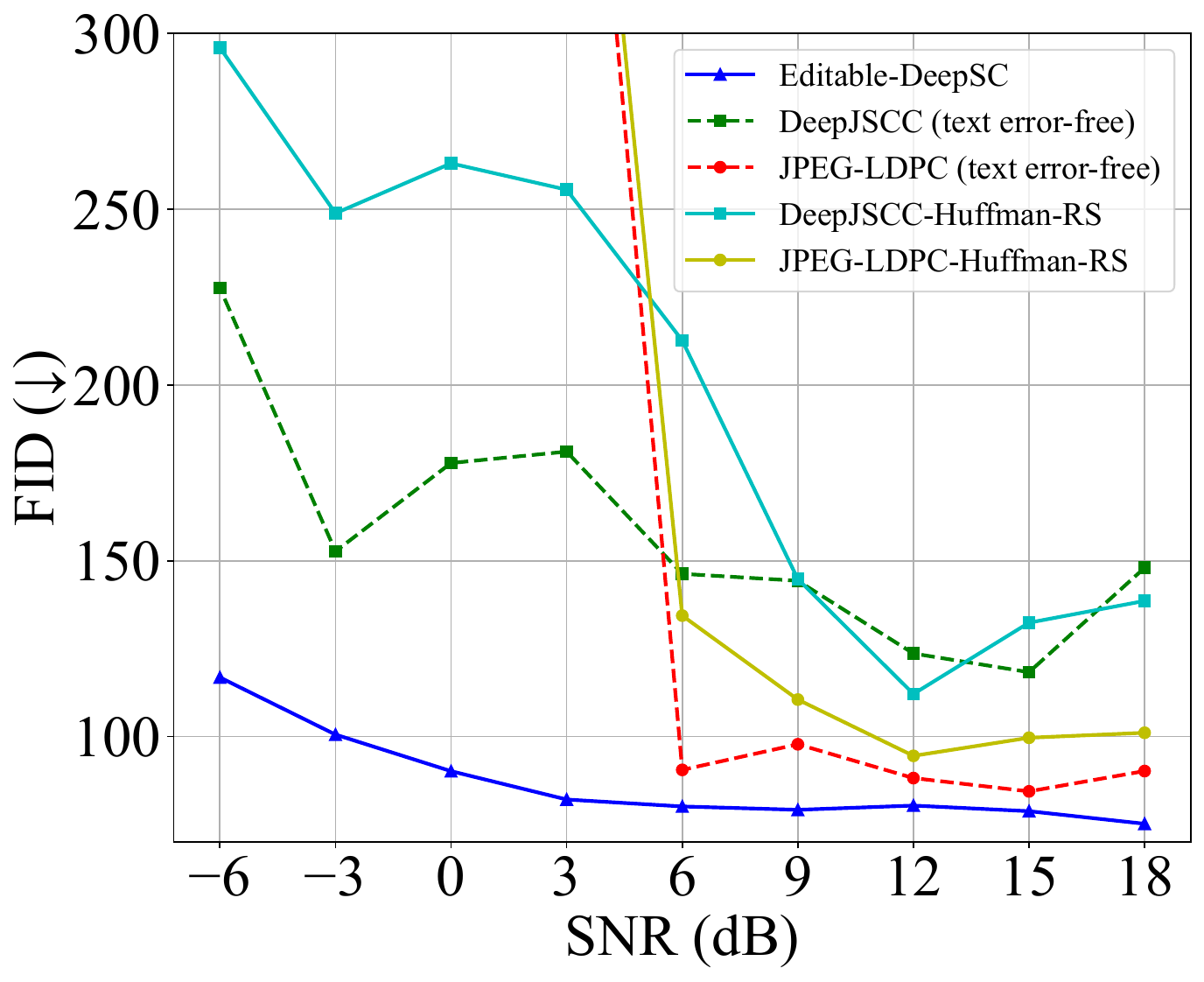}
        \label{fid}
    \end{subfigure}

    \vspace{-0.5cm}

    \caption{Quantitative comparison of different methods on cross-modal language-driven editing tasks. Note that $\uparrow$ indicates that the higher the better and $\downarrow$ indicates that the lower the better.}
    \label{quantitative}
\end{figure*}

To fairly evaluate the editing performance, we utilize several quantitative metrics, namely SSIM \cite{wang2004image}, LPIPS \cite{zhang2018unreasonable}, and FID \cite{heusel2017gans}. Higher SSIM values indicate better structural similarity between the edited images and the original images. Smaller LPIPS and FID values represent better perceptual preservation.

\section{Experiments}

We conduct the following experiments on CelebA-Dialog dataset \cite{jiang2021talk}, a famous visual-language facial editing dataset. We test the performance of our proposed \emph{Editable-DeepSC} on cross-modal language-driven editing tasks. We compare \emph{Editable-DeepSC} with classical \emph{data-oriented} communication methods. The \emph{data-oriented} approaches will respectively encode, transmit and recover the images and the corresponding textual instructions, after which the images and the texts will be encoded again and sent to the \emph{Semantic Editing Module} to acquire the edited images at the receiver side.

To evaluate the performance of \emph{data-oriented} methods on cross-modal language-driven editing tasks, we adopt the following transmission schemes:

\begin{itemize}
    \item \textbf{DeepJSCC (text error-free).} DeepJSCC \cite{bourtsoulatze2019deep} is utilized for image transfer. As for the text, we assume that the transfer is error-free, which means that the reconstructed sentences are identical to the original ones.
    \item \textbf{JPEG-LDPC (text error-free).} JPEG \cite{wallace1992jpeg} is utilized for image source coding and LDPC \cite{gallager1962low} is utilized for image channel coding. Text transfer is assumed to be error-free.
    \item \textbf{DeepJSCC-Huffman-RS.} DeepJSCC \cite{bourtsoulatze2019deep} is utilized for image transfer, Huffman \cite{huffman1952method} is utilized for text source coding, and RS \cite{reed1960polynomial} is utilized for text channel coding.
    \item \textbf{JPEG-LDPC-Huffman-RS.} JPEG \cite{wallace1992jpeg} is utilized for image source coding, LDPC \cite{gallager1962low} is utilized for image channel coding, Huffman \cite{huffman1952method} is utilized for text source coding, and RS \cite{reed1960polynomial} is utilized for text channel coding.
\end{itemize}

\begin{figure*}[htbp]
\centerline{\includegraphics[width=2\columnwidth]{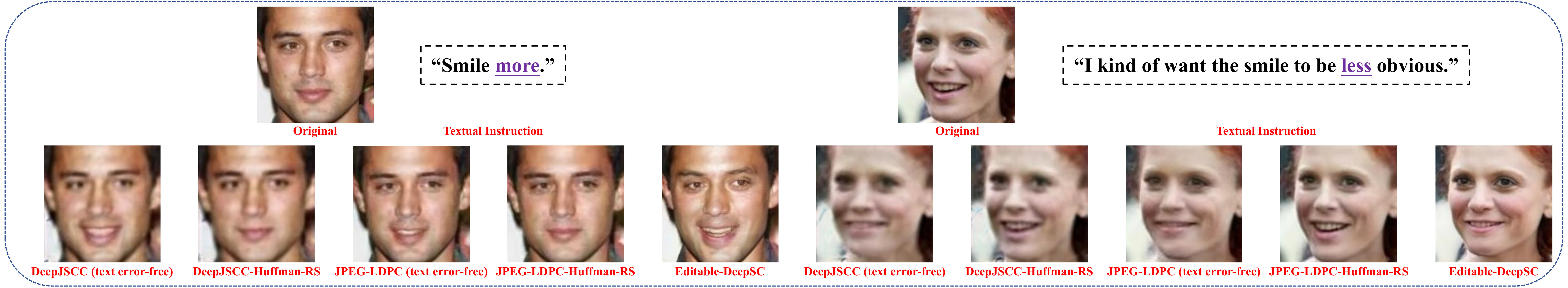}}
\caption{Qualitative comparison of different methods on cross-modal language-driven editing tasks ($6$ dB SNR). The original images and the textual instructions are presented in the $1$st row. The results of different methods are displayed in the $2$nd row.}
\label{qualitative}
\end{figure*}

The quantitative performance of different approaches on cross-modal language-driven editing tasks is presented in Figure \ref{quantitative}. We can conclude that \emph{Editable-DeepSC} achieves better SSIM performance than \emph{data-oriented} methods at almost all the circumstances of SNR, only worse than the approach \textbf{DeepJSCC (text error-free)} under the noise levels of $-6$ dB, $-3$ dB, and $0$ dB. However, the method \textbf{DeepJSCC (text error-free)} is simulated based on the assumptions that the text transmission is error-free, which rarely happens in the communications via noisy channels. As for the FID and LPIPS performance, \emph{Editable-DeepSC} behaves better than all the \emph{data-oriented} methods mentioned above under all the tested cases of SNR. These results demonstrate that \emph{Editable-DeepSC} can achieve remarkable editing effects in terms of fidelity and quality, outperforming \emph{data-oriented} methods. Besides, we observe that the performance of \textbf{JPEG-LDPC (text error-free)} and \textbf{JPEG-LDPC-Huffman-RS} decreases rapidly when the SNR is below $6$ dB, which is consistent with the fact that traditional communications suffer greatly from the \emph{cliff effect}.

Figure \ref{qualitative} illustrates the qualitative comparison between different communication approaches at the noise level of $6$ dB. \textbf{DeepJSCC-Huffman-RS} and \textbf{JPEG-LDPC-Huffman-RS} are unable to perform the anticipated editings because the instructive sentences have been severely damaged during the transmission process. Although \textbf{DeepJSCC (text error-free)} and \textbf{JPEG-LDPC (text error-free)} manage to change the smiles, their effects are not as vivid and natural as those achieved with \emph{Editable-DeepSC}.

\begin{table}[tbp]
\caption{Compression effectiveness of different communication methods for image transmission, measured by CBR.}
\begin{center}
\begin{tabular}{cc}
\toprule
Method & CBR ($\downarrow$) \\
\midrule
DeepJSCC & 0.083333 \\
JPEG-LDPC & 0.048709 \\
\textbf{Editable-DeepSC} & \textbf{0.010417} \\
\bottomrule
\end{tabular}
\end{center}
\label{tab:CBR}
\end{table}

Table \ref{tab:CBR} shows the compression effectiveness of different image transmission methods from the perspective of CBR defined in (\ref{CBR}). \emph{Editable-DeepSC} only utilizes around $12.5\%$ of \textbf{DeepJSCC} method's CBR and around $21.4\%$ of \textbf{JPEG-LDPC} method's CBR, yet it still achieves extraordinary editing effects and outperforms \emph{data-oriented} methods. \emph{Editable-DeepSC} not only performs satisfying editings, but can also considerably save the transmission bandwidth.

In summary, \emph{Editable-DeepSC} can achieve more realistic and satisfying editing effects than \emph{data-oriented} methods while significantly saving the communication overhead. This is because \emph{data-oriented} methods encode the original texts and images at the sender side, transmit them via the noisy channels, recover them at the receiver side, and encode them again so as to perform the pragmatic tasks at the receiver side, which will increase the data processing procedures and consequently result in information loss. The theory behind this is the well-known \textbf{Data Processing Inequality}\footnote{\cite{cover1991elements} \textbf{Theorem 2.8.1 (Data Processing Inequality)} If $X \rightarrow Y \rightarrow Z$, then $I(X;Y) \geq I(X;Z)$.} in information theory, which means that the semantic mutual information will decrease during the data processing procedures.

\section{Conclusion}

In this paper, we propose \emph{Editable-DeepSC}, a novel cross-modal editable semantic communication approach, to tackle the communication challenge under dynamic editing scenarios in a conversational and interactive way. We leverage the GAN inversion methods based on StyleGAN priors to extract the image features in a disentangled way. We further introduce the \emph{Semantic Editing Module} to perform fine-grained editings by iteratively updating the latent codes under the guidance of textual instructions. Extensive simulation results prove that \emph{Editable-DeepSC} can achieve superior performance compared to \emph{data-oriented} methods in terms of editing effects and transmission efficiency.

\section*{Acknowledgements}
This work is supported in part by the National Natural Science Foundation of China under grant 62171248, 62301189, Guangdong Basic and Applied Basic Research Foundation under grant 2021A1515110066, Guangdong Provincial Key Laboratory of Novel Security Intelligence Technologies (2022B1212010005), the PCNL KEY project (PCL2021A07), and Shenzhen Science and Technology Program under Grant JCYJ20220818101012025, RCBS20221008093124061, GXWD20220811172936001.

\bibliographystyle{IEEEtran}
\bibliography{references}

\begin{thebibliography}{10}
\providecommand{\url}[1]{#1}
\csname url@samestyle\endcsname
\providecommand{\newblock}{\relax}
\providecommand{\bibinfo}[2]{#2}
\providecommand{\BIBentrySTDinterwordspacing}{\spaceskip=0pt\relax}
\providecommand{\BIBentryALTinterwordstretchfactor}{4}
\providecommand{\BIBentryALTinterwordspacing}{\spaceskip=\fontdimen2\font plus
\BIBentryALTinterwordstretchfactor\fontdimen3\font minus
  \fontdimen4\font\relax}
\providecommand{\BIBforeignlanguage}[2]{{%
\expandafter\ifx\csname l@#1\endcsname\relax
\typeout{** WARNING: IEEEtran.bst: No hyphenation pattern has been}%
\typeout{** loaded for the language `#1'. Using the pattern for}%
\typeout{** the default language instead.}%
\else
\language=\csname l@#1\endcsname
\fi
#2}}
\providecommand{\BIBdecl}{\relax}
\BIBdecl

\bibitem{shannon1948mathematical}
C.~E. Shannon, ``A mathematical theory of communication,'' \emph{The Bell
  system technical journal}, vol.~27, no.~3, pp. 379--423, 1948.

\bibitem{weaver1953recent}
W.~Weaver, ``Recent contributions to the mathematical theory of
  communication,'' \emph{ETC: a review of general semantics}, pp. 261--281,
  1953.

\bibitem{wallace1992jpeg}
G.~K. Wallace, ``The jpeg still picture compression standard,'' \emph{IEEE
  transactions on consumer electronics}, vol.~38, no.~1, pp. xviii--xxxiv,
  1992.

\bibitem{huffman1952method}
D.~A. Huffman, ``A method for the construction of minimum-redundancy codes,''
  \emph{Proceedings of the IRE}, vol.~40, no.~9, pp. 1098--1101, 1952.

\bibitem{reed1960polynomial}
I.~S. Reed and G.~Solomon, ``Polynomial codes over certain finite fields,''
  \emph{Journal of the society for industrial and applied mathematics}, vol.~8,
  no.~2, pp. 300--304, 1960.

\bibitem{gallager1962low}
R.~Gallager, ``Low-density parity-check codes,'' \emph{IRE Transactions on
  information theory}, vol.~8, no.~1, pp. 21--28, 1962.

\bibitem{bourtsoulatze2019deep}
E.~Bourtsoulatze, D.~B. Kurka, and D.~G{\"u}nd{\"u}z, ``Deep joint
  source-channel coding for wireless image transmission,'' \emph{IEEE
  Transactions on Cognitive Communications and Networking}, vol.~5, no.~3, pp.
  567--579, 2019.

\bibitem{zhang2023adaptive}
Q.~Zhang, B.~Chen, Y.~Huang, and S.-T. Xia, ``Adaptive productae: Snr-aware
  adaptive decoding of neural product codes,'' in \emph{ICC 2023-IEEE
  International Conference on Communications}.\hskip 1em plus 0.5em minus
  0.4em\relax IEEE, 2023, pp. 6337--6342.

\bibitem{xie2021deep}
H.~Xie, Z.~Qin, G.~Y. Li, and B.-H. Juang, ``Deep learning enabled semantic
  communication systems,'' \emph{IEEE Transactions on Signal Processing},
  vol.~69, pp. 2663--2675, 2021.

\bibitem{xie2021task}
H.~Xie, Z.~Qin, and G.~Y. Li, ``Task-oriented multi-user semantic
  communications for vqa,'' \emph{IEEE Wireless Communications Letters},
  vol.~11, no.~3, pp. 553--557, 2021.

\bibitem{weng2023deep}
Z.~Weng, Z.~Qin, X.~Tao, C.~Pan, G.~Liu, and G.~Y. Li, ``Deep learning enabled
  semantic communications with speech recognition and synthesis,'' \emph{IEEE
  Transactions on Wireless Communications}, 2023.

\bibitem{goodfellow2020generative}
I.~Goodfellow, J.~Pouget-Abadie, M.~Mirza, B.~Xu, D.~Warde-Farley, S.~Ozair,
  A.~Courville, and Y.~Bengio, ``Generative adversarial networks,''
  \emph{Communications of the ACM}, vol.~63, no.~11, pp. 139--144, 2020.

\bibitem{xia2022gan}
W.~Xia, Y.~Zhang, Y.~Yang, J.-H. Xue, B.~Zhou, and M.-H. Yang, ``Gan inversion:
  A survey,'' \emph{IEEE Transactions on Pattern Analysis and Machine
  Intelligence}, vol.~45, no.~3, pp. 3121--3138, 2022.

\bibitem{fang2023gifd}
H.~Fang, B.~Chen, X.~Wang, Z.~Wang, and S.-T. Xia, ``Gifd: A generative
  gradient inversion method with feature domain optimization,'' in
  \emph{Proceedings of the IEEE/CVF International Conference on Computer
  Vision}, 2023, pp. 4967--4976.

\bibitem{fang2024privacy}
H.~Fang, Y.~Qiu, H.~Yu, W.~Yu, J.~Kong, B.~Chong, B.~Chen, X.~Wang, and S.-T.
  Xia, ``Privacy leakage on dnns: A survey of model inversion attacks and
  defenses,'' \emph{arXiv preprint arXiv:2402.04013}, 2024.

\bibitem{karras2019style}
T.~Karras, S.~Laine, and T.~Aila, ``A style-based generator architecture for
  generative adversarial networks,'' in \emph{Proceedings of the IEEE/CVF
  conference on computer vision and pattern recognition}, 2019, pp. 4401--4410.

\bibitem{zhang2018unreasonable}
R.~Zhang, P.~Isola, A.~A. Efros, E.~Shechtman, and O.~Wang, ``The unreasonable
  effectiveness of deep features as a perceptual metric,'' in \emph{Proceedings
  of the IEEE conference on computer vision and pattern recognition}, 2018, pp.
  586--595.

\bibitem{hochreiter1997long}
S.~Hochreiter and J.~Schmidhuber, ``Long short-term memory,'' \emph{Neural
  computation}, vol.~9, no.~8, pp. 1735--1780, 1997.

\bibitem{jiang2021talk}
Y.~Jiang, Z.~Huang, X.~Pan, C.~C. Loy, and Z.~Liu, ``Talk-to-edit: Fine-grained
  facial editing via dialog,'' in \emph{Proceedings of the IEEE/CVF
  International Conference on Computer Vision}, 2021, pp. 13\,799--13\,808.

\bibitem{deng2019arcface}
J.~Deng, J.~Guo, N.~Xue, and S.~Zafeiriou, ``Arcface: Additive angular margin
  loss for deep face recognition,'' in \emph{Proceedings of the IEEE/CVF
  conference on computer vision and pattern recognition}, 2019, pp. 4690--4699.

\bibitem{wang2004image}
Z.~Wang, A.~C. Bovik, H.~R. Sheikh, and E.~P. Simoncelli, ``Image quality
  assessment: from error visibility to structural similarity,'' \emph{IEEE
  transactions on image processing}, vol.~13, no.~4, pp. 600--612, 2004.

\bibitem{heusel2017gans}
M.~Heusel, H.~Ramsauer, T.~Unterthiner, B.~Nessler, and S.~Hochreiter, ``Gans
  trained by a two time-scale update rule converge to a local nash
  equilibrium,'' \emph{Advances in neural information processing systems},
  vol.~30, 2017.

\bibitem{cover1991elements}
T.~M. Cover and J.~A. Thomas, \emph{Elements of Information Theory}.\hskip 1em
  plus 0.5em minus 0.4em\relax John Wiley \& Sons, 1991.

\end{thebibliography}

\end{document}